\documentclass[pre,showpacs,preprint]{revtex4-1}
\usepackage{graphicx}
\usepackage{amssymb}
\usepackage{amsmath}
\usepackage{ulem}
\usepackage[usenames]{color}

\newcommand\st{\bgroup\markoverwith
{\textcolor{blue}{\rule[0.5ex]{2pt}{0.4pt}}}\ULon}

\begin{document}
\title{Vortex interaction on curved surfaces}
\author{Seung Ki Baek}
\affiliation{School of Physics, Korea Institute for Advanced Study, Seoul
130-722, Korea}
\email[Corresponding author, E-mail: ]{seungki@kias.re.kr}

\begin{abstract}
The vortex-excitation energy on a sphere can be obtained by using the
stereographic projection. By applying this method, we calculate the energy
needed to create a vortex on a surface with a constant negative curvature.
It is found that the energy is a linear function of the radius of the vortex.
In accordance with this result, the interaction energy between a pair of
vortices is also found to change linearly with the vortex separation distance.
Explicit vortex configurations are obtained numerically with this interaction.
\end{abstract}


\maketitle

\section{Introduction}
There is growing interest in topological defects on curved surfaces.
One classical example is the Thomson problem, which addresses the question
of how to
configure charges on a sphere with minimal energy~\cite{thomson}. A
spherical virus cell having subcellular structures on its surface can be
viewed as a biological counterpart of the Thomson problem. The orientational
order of liquid crystal molecules in a curved sheet provides another example
of the interaction between defects and curvature~\cite{park}. This system can
be formulated as an $XY$ spin model where the curvature term enters the
Hamiltonian in a very similar way to that of the magnetic vector potential
in the theory of type-II superconductors (see, for detailed
discussions, Ref.~\cite{turner} and references therein).
Owing to the obvious ubiquity of spherical shapes, the physics of defects on
a surface with positive Gaussian curvature is relatively well
understood~\cite{shin,*xing,*zhang} and a number of experiments have been
performed to check the theoretical
understanding~\cite{nieves,*lopez,*leon,*xing2,*jeong}.
 
We have been interested in $XY$-type
models on a surface with negative curvature~\cite{xyhep,*geomxy,qclock}.
Hyperbolic geometry on such a surface is also an important model of
non-Euclidean geometry~\cite{anderson}, and in physical contexts,
a negatively curved surface has been introduced as a conceptual tool
to understand disordered systems without intrinsic
randomness~\cite{auriac,sausset08}.
To our knowledge, however, it is not
entirely clear how the interaction between topological defects, or vortices,
depends on distance in the case of negatively curved surfaces. For example, the
potential was predicted to be short ranged in Ref.~\cite{belo} while the
generalized Gauss law predicts it to be very long ranged~\cite{turner}.
Roughly speaking, the main difference between these two alternatives can be
traced to whether or not the curvature appears as a source term in Gauss's
law:
\begin{equation}
\oint_{\partial V} \mathbf{E} \cdot d\mathbf{S} = \int_V \sigma~ dV,
\label{eq:gauss}
\end{equation}
where $V$ is a volume enclosed by a boundary surface $\partial V$.
Gauss's law states that the surface integral of field $\mathbf{E}$ over
$\partial V$ should match with the volume integral of source
terms distributed with density $\sigma$.
An important fact is that a circle on a negatively curved surface has
an exponentially growing boundary as the radius increases. As a consequence,
if only a fixed amount of defects contributes to the source term, the field
strength should decay exponentially. If the curvature also serves as a
source term, however, Eq.~(\ref{eq:gauss}) describes competition between the
surface and volume, both of which increase in the same exponential manner
as the length scale of $V$ grows, so the field strength never vanishes
no matter how far away one gets from the vortex core.
Although theoretical predictions about the $XY$ model on a curved surface
are mainly based on the former scenario~\cite{belo,auriac}, a numerical
study suggests that a finite temperature is needed to unbind
vortex-antivortex pairs on a curved surface~\cite{xyhep,geomxy}.
It implies that energy should be able to compete with entropy in creating
vortices, which will increase logarithmically with the volume of the system,
meaning that the vortex interaction cannot decay so fast with distance.

In this work, we directly calculate the excitation energy using the
stereographic projection method and verify that the latter case is the
correct alternative.
This work is organized as follows: Sec.~\ref{sec:sphere} reviews the
stereographic projection applied to the spherical case, which is intended to
be a mild introduction to the basic formalism. We then proceed to the case
of negative curvature in Sec.~\ref{sec:pseudo}, where we also present
numerically obtained vortex configurations based on the potential form. We
conclude this work in Sec.~\ref{sec:summary}.

\section{Unit sphere}
\label{sec:sphere}

\begin{figure}
\begin{center}
\includegraphics[width=0.45\textwidth]{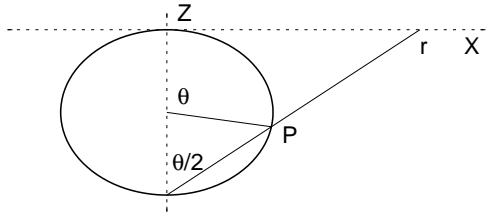}
\end{center}
\caption{Stereographic projection of a unit sphere onto a plane. A point
$\mathbf{P}$ on a sphere, specified by $(\theta,\phi)$, is projected onto
$z=r e^{i\phi}$ with $r=2\tan \frac{\theta}{2}$.}
\label{fig:stereo}
\end{figure}

We start with reviewing how defects interact on a unit sphere, i.e., a
surface with a constant positive curvature. The outline of this calculation
has already been explained in Ref.~\cite{luben} and the purpose of this section
is to present the general method before proceeding to the case of negative
curvature.

Let us define coordinates $(u,v) = (\theta,\phi)$ and
$(u',v') = (z,\bar{z})$ related by the stereographic projection. As clearly
seen in Fig.~\ref{fig:stereo}, the projection maps the original spherical
coordinate $(\theta,\phi)$ onto a complex variable $z=2\tan
\frac{\theta}{2}~e^{i \phi}$ and its complex conjugate $\bar{z}=2\tan
\frac{\theta}{2}~e^{-i\phi}$.
In this spherical coordinate, a point on the sphere is expressed as
$\mathbf{P} = (X,Y,Z) = (\sin\theta \cos\phi, \sin\theta \sin\phi, \cos\theta)$
and the length of a line element, $ds$, is given by
$ds^2 = g_{00}~d\theta^2 + (g_{01} + g_{10}) ~d\theta d\phi + g_{11} ~d\phi^2
= d\theta^2 + \sin^2\theta~d\phi^2$.
This defines the metric tensor as
\begin{equation*}
\tilde{g}
= \left( \begin{array}{cc} g_{00} & g_{01}\\ g_{10} & g_{11}\end{array} \right)
= \left( \begin{array}{cc} 1 & 0\\ 0 & \sin^2 \theta \end{array} \right).
\end{equation*}
On the other hand, we want to work with the other coordinate system
$(u',v')$ and the metric tensor $g'$ in the new coordinate $(u',v')$ is
transformed to $g$ by
$\tilde{g} = U^{T} \cdot \tilde{g}' \cdot U$
with the Jacobian matrix
\begin{equation*}
U \equiv
\left( \begin{array}{cc}
\frac{\partial u'}{\partial u} & \frac{\partial u'}{\partial v}\\
\frac{\partial v'}{\partial u} & \frac{\partial v'}{\partial v}
\end{array} \right)
= \left( \begin{array}{cc}
\sec^2 \frac{\theta}{2}~e^{i\phi} & 2i\tan \frac{\theta}{2}~e^{i\phi}\\
\sec^2 \frac{\theta}{2}~e^{-i\phi} & -2i\tan \frac{\theta}{2}~e^{-i\phi}
\end{array} \right).
\end{equation*}
It is now straightforward to obtain $g'$ using the inverse matrix
\begin{equation*}
U^{-1} =
\left( \begin{array}{cc}
\frac{1}{2} \cos^2 \frac{\theta}{2}~e^{-i\phi} &
\frac{1}{2} \cos^2 \frac{\theta}{2}~e^{i\phi} \\
\frac{1}{4i} \cot \frac{\theta}{2}~e^{-i\phi} &
-\frac{1}{4i} \cot \frac{\theta}{2}~e^{i\phi}
\end{array} \right),
\end{equation*}
resulting in
\begin{eqnarray}
\tilde{g}' &=& \left( U^{-1} \right)^{T} \cdot \tilde{g} \cdot U^{-1}
=\frac{1}{2} \cos^4 \frac{\theta}{2}
\left( \begin{array}{cc} 0 & 1 \\ 1 & 0 \end{array} \right)\nonumber\\
&=&\frac{1}{2 \left(1+z\bar{z}/4\right)^2}
\left( \begin{array}{cc} 0 & 1 \\ 1 & 0 \end{array} \right)
= \left( \begin{array}{cc} g'_{00} & g'_{01} \\ g'_{10} & g'_{11}
\end{array} \right)
= \left(g_{z\bar{z}}\right).
\label{eq:g'}
\end{eqnarray}
Note that the equality between the first and second lines follows from
$z\bar{z} = 4\tan^2 \frac{\theta}{2} = 4 \left( \sec^2 \frac{\theta}{2}-1
\right)$.
Equation~(\ref{eq:g'}) means that a line element on the projected plane will
be expressed as
\begin{eqnarray}
ds^2 &=&
\left( \begin{array}{cc} dz & d\bar{z} \end{array} \right)
\left( \begin{array}{cc} g'_{00} & g'_{01} \\ g'_{10} & g'_{11} \end{array}
\right)
\left( \begin{array}{c} dz \\ d\bar{z} \end{array} \right)\nonumber\\
&=& \frac{1}{\left[ 1+(x^2+y^2)/4 \right]^2} \left( dx^2 + dy^2 \right),
\label{eq:smetric}
\end{eqnarray}
if one writes $z=x+iy$. Hence, we can say that the determinant of the metric
tensor in the $(x,y)$ plane is
$g \equiv \det \tilde{g} = \left[ 1+(x^2+y^2)/4 \right]^{-4}$.
Integrating the area over the whole complex plane therefore
yields
$\int dS = \iint \sqrt{g}~dx~dy = \int_0^{\infty} 2 \pi r \left(
1+r^2/4 \right)^{-2} dr = 4\pi$,
which is exactly the surface area of the unit sphere. Also note that the
inverse metric tensor is given by
\begin{equation*}
\tilde{g}^{-1} = \left( \begin{array}{cc} g^{00} & g^{01} \\ g^{10} & g^{11}
\end{array} \right) = 2(1+z\bar{z}/4)^2
\left( \begin{array}{cc} 0 & 1 \\ 1 & 0 \end{array} \right),
\end{equation*}
where we omit the prime ($'$) to indicate $(u',v')$ for brevity.
When $ds = w(z) |dz|$, the Gaussian curvature $K$ is given
by the following formula~\cite{anderson}:
\begin{eqnarray}
K &=& - \frac{4}{w^2(z)} \left[
\frac{\partial^2}{\partial z \partial \bar{z}} \ln w(z) \right]\nonumber\\
&=&  - \frac{4}{w^2(x,y)}
\left[ \frac{1}{2} \left( \frac{\partial}{\partial x} -i
\frac{\partial}{\partial y} \right) \right]
\left[ \frac{1}{2} \left( \frac{\partial}{\partial x} +i
\frac{\partial}{\partial y} \right) \right] \ln w(x,y)\nonumber\\
&=&  - \frac{1}{w^2(x,y)} \left( \frac{\partial^2}{\partial x^2} +
\frac{\partial^2}{\partial y^2} \right) \ln w(x,y).
\label{eq:curvature}
\end{eqnarray}
By substituting Eq.~(\ref{eq:smetric}) here, we find $K = 1$ for the
unit sphere as expected.

We have so far studied the fundamental property of the surface.
The next step is to
place physical objects on it. Let us consider a vector field
$\mathbf{m}(\mathbf{x}) = \cos \gamma(\mathbf{x}) ~\mathbf{e}_1(\mathbf{x})
+ \sin \gamma(\mathbf{x}) ~\mathbf{e}_2 (\mathbf{x})$ on the surface.
In tracing out its changes, however, it should be taken into account
that the coordinate system $\left( \mathbf{e}_1(\mathbf{x}),
\mathbf{e}_2(\mathbf{x})\right)$ itself depends on the position
$\mathbf{x}$. A new vector field called the connection, or spin
connection, enters here, which is derived from the given coordinate system by
$\mathbf{A}(\mathbf{x}) = \mathbf{e}_1(\mathbf{x}) \cdot \nabla
\mathbf{e}_2 (\mathbf{x})$~\cite{kamien}.
On the unit sphere, the most natural coordinate system would be obtained by
differentiating $\mathbf{P}$, i.e., 
$\mathbf{e}_\theta = \left( \cos\theta \cos\phi, \cos\theta \sin\phi,
-\sin\theta \right)$ and
$\mathbf{e}_\phi = \left( -\sin\phi, \cos\phi, 0 \right)$.
We express the connection in the $(z,\bar{z})$ coordinate as
$(A_z, A_{\bar{z}}) = \left(\mathbf{e}_\theta \cdot \partial \mathbf{e}_\phi
/ \partial z, ~\mathbf{e}_\theta \cdot
\partial \mathbf{e}_\phi / \partial \bar{z} \right)$, and
it is fairly straightforward to see that
$A_z = \mathbf{e}_\theta \cdot \partial \mathbf{e}_\phi / \partial z 
= -\cos\theta \left( \partial \phi / \partial z \right)$. The cosine part
can be easily expressed in the new coordinate system, since
$\cos\theta = 2\cos^2 \frac{\theta}{2} - 1 =
\frac{1-z\bar{z}/4}{1+z\bar{z}/4}$.
And it follows from $z = r e^{i\phi}$ that
$\phi = \mbox{Im} \ln z = \frac{1}{2i} \left( \ln z - \ln \bar{z} \right)$,
so we find that
$\partial \phi / \partial z = (2iz)^{-1}$ and
$\partial \phi / \partial \bar{z} = -(2i\bar{z})^{-1}$.
In short, we obtain the connection as
\begin{equation}
A_z = -\frac{1}{2iz} \left( \frac{1-z\bar{z}/4}{1+z\bar{z}/4} \right) =
\bar{A}_{\bar{z}}.
\label{eq:con1}
\end{equation}
Let us now consider the contribution to the free energy due to the
curvature of the surface, which is usually called the Frank free energy.
In the one-constant approximation, that is, if three elastic constants
associated with splay, twist, and bend are of an equal size~\cite{liquid},
the Frank free energy assumes the following form:
\begin{equation}
F = \frac{K_A}{2} \iint dx_{\alpha} dx_{\beta} \sqrt{g}~ g^{\alpha \beta}
\left( \frac{\partial \gamma}{\partial x_{\alpha}} - A_{\alpha} \right) \left(
\frac{\partial \gamma}{\partial x_{\beta}} - A_{\beta} \right),
\label{eq:f}
\end{equation}
with the Frank constant $K_A$.
On a flat surface, the connection $\mathbf{A}$ can be set as zero
and the remaining part describes the usual Goldstone mode.
In Eq.~(\ref{eq:f}),
putting $\gamma=0$ automatically introduces
one defect at $\theta=0$ and another at $\theta=\pi$.
This is argued in Ref.~\cite{luben} by pointing out that
\begin{equation*}
A_z \longrightarrow \left\{
\begin{array}{ll}
-\frac{1}{2iz} & \mbox{~~~if~~} z \rightarrow 0,\\
+\frac{1}{2iz} & \mbox{~~~if~~} z \rightarrow \infty.
\end{array}
\right.
\end{equation*}
The Poincar\'{e}-Brouwer theorem dictates that the sum
of defects on a closed surface should be equal to the Euler characteristic
$\chi$ by
\begin{equation}
\chi = \frac{1}{2\pi} \int K dS = \frac{1}{2\pi} \oint \left[
\nabla \times \nabla \theta (\mathbf{x}) \right] \cdot d \mathbf{S},
\label{eq:pb}
\end{equation}
since $\nabla \times \nabla \theta(\mathbf{x}) = m\delta^2(\mathbf{x})$ for
a defect with charge $m$~\cite{kamien}. The Euler characteristic is related
to the genus $g$ of the surface, i.e., the number of handles, by $\chi = 2 -
2g$. A sphere therefore has $\chi=2$,
which is of course consistent with the two defects that we have now, and the
vector field given above connects these defects by geodesics. Additional
discussion on Eq.~(\ref{eq:f}) and two-dimensional electrostatics is
presented in Appendix~\ref{app:elec}.
Since $A_z A_{\bar{z}} = \frac{1}{4z\bar{z}} \left(
\frac{1-z\bar{z}/4}{1+z\bar{z}/4} \right)^2$,
the calculation reduces to
\begin{eqnarray*}
F &=& \frac{K_A}{2} \iint \sqrt{g}~ g^{\alpha \beta}
A_{\alpha} A_{\beta} ~dx_{\alpha} ~dx_{\beta} \nonumber\\
&=& \frac{K_A}{2} \int \frac{1}{(1+r^2/4)^2} \times 4 \left( 1+r^2/4
\right)^2 \times \frac{1}{4r^2} \left( \frac{1-r^2/4}{1+r^2/4}
\right)^2 ~2\pi r ~dr\nonumber\\
&=& \pi K_A \int \frac{1}{r} \left( \frac{1-r^2/4}{1+r^2/4} \right)^2 dr.
\end{eqnarray*}
Let us compute the energy inside a hemisphere, which corresponds to $r <
2\tan \frac{\pi}{4} = 2$ (see Fig.~\ref{fig:stereo}), assuming that
a defect has a very small core radius, $\epsilon \ll 1$.
The answer is
\begin{equation*}
\frac{F}{2} = \pi K_A \int_\epsilon^2 \frac{1}{r} \left( \frac{1-r^2/4}{1+r^2/4}
\right)^2 dr \approx \pi K_A \left( \ln \frac{2}{\epsilon} - 1 \right),
\end{equation*}
and thus the Frank free energy over the whole sphere is given as
$F = 2 \pi K_A \left( \ln \frac{2}{\epsilon} - 1 \right)$.
Alternatively, one may carry out the integration
over $\left[ \epsilon,4/\epsilon \right]$ as
\begin{equation*}
F = \pi K_A \int_\epsilon^{4/\epsilon}
\frac{1}{r} \left( \frac{1-r^2/4}{1+r^2/4}
\right)^2 dr \approx 2\pi K_A \left( \ln \frac{2}{\epsilon} - 1 \right),
\end{equation*}
since the boundary of the defect at $\theta = \pi$ is projected onto a
circle of radius $4/\epsilon$ on the complex plane.
Having dealt with a highly symmetric configuration as above, we
may try a slightly more general case. We replace the defect at $z=0$ by
another one at $z=z_0$ by setting $\gamma = -\mbox{Im} \ln z + \mbox{Im}
\ln(z-z_0)$ and this leads to
\begin{eqnarray*}
\frac{\partial \gamma}{\partial z} - A_z
&=& -\frac{1}{2iz} + \frac{1}{2i(z-z_0)} +
\frac{1}{2iz}\left(\frac{1-z\bar{z}/4}{1+z\bar{z}/4}
\right)\nonumber\\
&=& \frac{1}{2i(z-z_0)} -
\frac{1}{2iz} \left( \frac{2 z\bar{z}/4}{1+z\bar{z}/4} \right),
\end{eqnarray*}
and
\begin{equation*}
\left| \frac{\partial \gamma}{\partial z} - A_z \right|^2 = \frac{1}{4
|z-z_0|^2} + \frac{|z|^2/4}{4\left( 1+|z|^2/4 \right)^2} -
\frac{|z|^2 - (z \bar{z}_0 + \bar{z} z_0)/2}{4|z-z_0|^2 \left(1+|z|^2/4
\right)}.
\end{equation*}
Let us denote the distance of this defect from the origin of the
complex plane as $|z_0| \equiv \rho$.
From Fig.~\ref{fig:stereo}, we see that the image of the defect at
$\theta$ ranges over $2\tan\frac{\theta \pm \epsilon}{2} \approx 2 \tan
\frac{\theta}{2} \pm \epsilon \sec^2 \frac{\theta}{2} = \rho \pm
\epsilon \left( 1+\rho^2/4 \right)$ by a simple expansion and therefore the
projected defect will have a radius of $\epsilon' \equiv \epsilon
(1+\rho^2/4)$. We calculate the corresponding Frank free energy
\begin{eqnarray}
F &=& \frac{K_A}{2} \iint \frac{1}{|z-z_0|^2} ~dz ~d\bar{z} + \frac{K_A}{8}
\iint \frac{|z|^2}{(1+|z|^2/4)^2} ~dz ~d\bar{z}\nonumber\\
&-& \frac{K_A}{2} \iint \frac{|z|^2 - (z \bar{z}_0 + \bar{z} z_0)/2}{
|z-z_0|^2 \left(1+|z|^2/4\right)} ~dz ~d\bar{z},
\label{eq:frank}
\end{eqnarray}
and the first term is evaluated as $\pi K_A \ln
\frac{4}{\epsilon \epsilon'}$ when $\rho \sim
O(1)$ (Appendix~\ref{app:term1}). The second term is obtained as
$\frac{K_A}{8} \int_{\epsilon}^{4/\epsilon} r^2 (1+r^2/4)^{-2} ~2\pi r
~dr \approx 2\pi K_A \left( 2\ln\frac{2}{\epsilon} - 1\right)$,
and the last one is $-2 \pi K_A \ln \frac{4}{\epsilon \epsilon'}$
(Appendix~\ref{app:term3}). Summing them up, we find that
\begin{equation*}
F = 2\pi K_A \left( \ln \frac{2}{\epsilon}\sqrt{1+\rho^2/4} - 1 \right) = 2
\pi K_A \left( \ln \frac{2}{\epsilon \cos\frac{\theta}{2}} -1 \right).
\end{equation*}
In the case where two vortices are present, their configuration with minimizing
the energy is therefore found at $\rho=0$, where they are located at
exactly opposite sides of the sphere.

\section{Unit Pseudosphere}
\label{sec:pseudo}

\begin{figure}
\begin{center}
\includegraphics[width=0.45\textwidth]{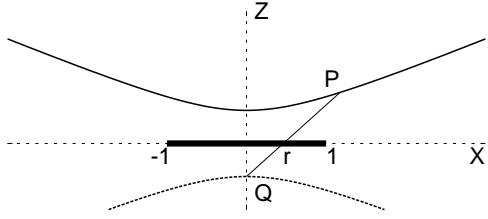}
\end{center}
\caption{Projection of a unit hyperboloid onto the Poincar\'e disk on the
complex plane. The disk is represented by the thick solid line at $Z=0$.}
\label{fig:pseudo}
\end{figure}

In order to deal with a surface having negative curvature,
first we draw a hyperboloid by rotating a hyperbola around its semi-major
axis. The three-dimensional shape is described by $X^2 + Y^2 - Z^2 = -1$.
After parameterizing a point on the hyperboloid by
$\mathbf{P} = (X,Y,Z) = (\sinh\theta \cos\phi, \sinh\theta \sin\phi,
\cosh\theta)$,
one may consider a projection onto a plane, $Z=0$, in viewing the
hyperboloid from $\mathbf{Q}=(0,0,-1)$ (Fig.~\ref{fig:pseudo}). By simple
algebra, we find that the resulting point on the plane can be written as
$(x,y) = (r \cos\phi, r \sin\phi)$ with $r=\tanh \frac{\theta}{2}$.
Therefore, we have two coordinate systems, i.e.,
$(u,v)=(\theta,\phi)$ and $(u',v') = (z, \bar{z}) = (\tanh
\frac{\theta}{2}~e^{i\phi}, \tanh \frac{\theta}{2}~e^{-i\phi})$, and the
unit disk on the complex plane covered by this projection is called the
Poincar\'e disk.
We want this hyperboloid to be considered as a sort of {\em sphere}, so we
define a dot product between two vectors $\mathbf{I}=(I_1,
I_2, I_3)$ and $\mathbf{J}=(J_1, J_2, J_3)$ as
$\mathbf{I} \cdot \mathbf{J} \equiv I_1 J_1 + I_2 J_2 - I_3 J_3$.
We can then simply describe the hyperboloid by
$\mathbf{P} \cdot \mathbf{P} = -1$.
According to this dot product, the line element in terms of $(\theta,\phi)$
is given by
\begin{equation}
ds^2 = \left( \frac{\partial \mathbf{P}}{\partial \theta} \cdot \frac{\partial
\mathbf{P}}{\partial \theta} \right) d^2\theta
+ \left( \frac{\partial \mathbf{P}}{\partial
\phi} \cdot \frac{\partial \mathbf{P}}{\partial \phi} \right) d^2\phi
= d^2 \theta + \sinh^2 \theta ~d^2 \phi,
\label{eq:metric2}
\end{equation}
which defines the metric tensor as
\begin{equation*}
\tilde{g}
= \left( \begin{array}{cc} g_{00} & g_{01} \\ g_{10} & g_{11} \end{array}
\right)
= \left( \begin{array}{cc} 1 & 0 \\ 0 & \sinh^2 \theta \end{array} \right).
\end{equation*}
Note from Eq.~(\ref{eq:metric2}) that $\theta$ directly represents the
radial distance in this metric.
By using the Jacobian matrix
\begin{equation*}
U \equiv
\left( \begin{array}{cc}
\frac{\partial u'}{\partial u} & \frac{\partial u'}{\partial v}\\
\frac{\partial v'}{\partial u} & \frac{\partial v'}{\partial v}
\end{array} \right)
= \left( \begin{array}{cc}
\frac{1}{2} \cosh^{-2} \frac{\theta}{2}~e^{i\phi} & i\tanh \frac{\theta}{2}~e^{i\phi}\\
\frac{1}{2} \cosh^{-2} \frac{\theta}{2}~e^{-i\phi} & -i\tanh \frac{\theta}{2}~e^{-i\phi}
\end{array} \right),
\end{equation*}
one can express the metric tensor in the $(u',v')$ coordinate as
\begin{eqnarray}
\tilde{g}' &=& \left( U^{-1} \right)^{T} \cdot \tilde{g} \cdot U^{-1}
=2 \cosh^4 \frac{\theta}{2}
\left( \begin{array}{cc} 0 & 1 \\ 1 & 0 \end{array} \right)\nonumber\\
&=&\frac{2}{\left(1-z\bar{z}\right)^2}
\left( \begin{array}{cc} 0 & 1 \\ 1 & 0 \end{array} \right)
= \left( \begin{array}{cc} g'_{00} & g'_{01} \\ g'_{10} & g'_{11}
\end{array} \right)
= \left( g_{z \bar{z}} \right),
\end{eqnarray}
since $z\bar{z} = \tanh^2 \frac{\theta}{2} = 1 - \cosh^{-2}
\frac{\theta}{2}$. If one writes $z=x+iy$, this new metric means that
\begin{eqnarray}
ds^2 &=&
\left( \begin{array}{cc} dz & d\bar{z} \end{array} \right)
\left( \begin{array}{cc} g'_{00} & g'_{01} \\ g'_{10} & g'_{11} \end{array}
\right)
\left( \begin{array}{c} dz \\ d\bar{z} \end{array} \right)\nonumber\\
&=& \frac{4}{\left[ 1-(x^2+y^2) \right]^2} \left( dx^2 + dy^2 \right),
\label{eq:hmetric}
\end{eqnarray}
so we find that
$\sqrt{g} = 4\left[ 1-(x^2+y^2) \right]^{-2}$ on the complex plane.
By using Eq.~(\ref{eq:curvature}) together with Eq.~(\ref{eq:hmetric}), one
can readily confirm that $K = -1$.
For a circle of radius $R = \tanh \frac{\theta^{\ast}}{2}$ on the projected
plane, the area inside the circle will be
$\int dS = \iint \sqrt{g}~ dx~dy =
\int_0^R 4\left( 1-r^2 \right)^{-2} 2 \pi r~dr = 4\pi
R^2 (1-R^2)^{-1} = 4\pi \sinh^2 \frac{\theta^{\ast}}{2}$,
while it is simply $\pi {\theta^\ast}^2$ on the Euclidean plane.

By differentiating the position vector $\mathbf{P}$, we obtain basic unit
vectors to define a coordinate system on this surface, i.e., 
$\mathbf{e}_\theta = \left( \cosh\theta \cos\phi, \cosh\theta \sin\phi,
\sinh\theta \right)$ and
$\mathbf{e}_\phi = \left( -\sin\phi, \cos \phi, 0 \right)$.
Since
$A_z = \mathbf{e}_\theta \cdot \partial \mathbf{e}_\phi / \partial z
= -\cosh\theta~ \frac{\partial \phi}{\partial z}$ and
$\cosh\theta = 2\cosh^2\frac{\theta}{2}-1 =
\frac{1+z\bar{z}}{1-z\bar{z}}$,
the connection is obtained as
\begin{equation}
A_z = -\frac{1}{2iz} \left( \frac{1+z\bar{z}}{1-z\bar{z}} \right) =
\bar{A}_{\bar{z}}.
\label{eq:con2}
\end{equation}
We again consider a vector field $\mathbf{m}(\mathbf{x}) = \cos
\gamma(\mathbf{x}) ~\mathbf{e}_\theta(\mathbf{x}) + \sin \gamma(\mathbf{x})
~\mathbf{e}_\phi (\mathbf{x})$. It is notable that $\mathbf{m} \cdot
\mathbf{m} = 1$ is satisfied at any $\mathbf{x}$ with our new dot product as
well. The simplest possible configuration would be to
set $\gamma = 0$ to introduce a defect at
$(X,Y,Z)=(0,0,1)$. The Frank free energy is then written as
\begin{eqnarray*}
F &=& \frac{K_A}{2} \iint \sqrt{g}~ g^{\alpha \beta}~ A_{\alpha} A_{\beta}
~dx_{\alpha} ~dx_{\beta}\nonumber\\
&=& \frac{K_A}{2} \int \frac{4}{\left( 1-r^2 \right)^2} \times \left( 1-r^2
\right)^2 \times \frac{1}{4r^2} \left( \frac{1+r^2}{1-r^2} \right)^2 ~2\pi
r~dr.
\end{eqnarray*}
Assuming a very small defect core radius $\epsilon \ll 1$,
the integration up to $R = \tanh \frac{\theta^{\ast}}{2}$ yields
\begin{equation*}
F \approx \pi K_A \left( \frac{2R^2}{1-R^2} + \ln \frac{R}{\epsilon} \right)
= \pi K_A \left( 2 \sinh^2{\frac{\theta^{\ast}}{2}} + \ln
\frac{\tanh\frac{\theta^{\ast}}{2}}{\epsilon} \right).
\end{equation*}
Note that the first term is proportional to the total area and the second
term corresponds to the Coulomb potential on the pseudosphere~\cite{jan}:
\begin{equation}
F_C = \pi K_A \ln \tanh \frac{\theta^{\ast}}{2}.
\label{eq:coulomb}
\end{equation}
In fact, we can remove the vortex at the center by assuming $\gamma =
-\mbox{Im} \ln z$
since there is no such restriction as Eq.~(\ref{eq:pb}) on the total
sum of defects on this surface which is not closed.
In this case, we observe
\begin{eqnarray*}
F_0 &=& \frac{K_A}{2} \int_0^R 4 \frac{r^2}{(1-r^2)^2} 2\pi r~dr = 2\pi K_A
\left[ \frac{R^2}{1-R^2} + \ln(1-R^2) \right]\nonumber\\
&\approx& 2\pi K_A \left(
\sinh^2 \frac{\theta^{\ast}}{2} -\theta^\ast \right). 
\end{eqnarray*}
The net contribution from creating the vortex is therefore
\begin{equation}
\Delta F = F - F_0 \approx \pi K_A \left( \ln \frac{\tanh
\frac{\theta^{\ast}}{2}}{\epsilon} + 2 \theta^\ast \right).
\label{eq:df}
\end{equation}
Since the magnitude of $\ln \tanh x$ becomes very small as $x$ increases,
the above expression is approximated as $\Delta F \approx 2\pi K_A
\theta^\ast$ at any moderate distance. Interestingly, it is a logarithmic
function of the area occupied by the vortex as in the planar case.
It is reasonable to guess that the same functional form as in
Eq.~(\ref{eq:df}) describes the interaction potential $E$ between two
vortices separated by a hyperbolic distance $\theta^\ast$,
\begin{equation}
E(\theta^\ast) \approx -J q_1 q_2 \theta^\ast,
\label{eq:u}
\end{equation}
where $J \approx 2 \pi K_A$ means strength of the interaction, and $q_1$ and
$q_2$ mean charges of the two vortices, respectively. The overall sign in
Eq.~(\ref{eq:u}) is chosen so that a vortex repels (attracts) another vortex
with the same (different) sign. If we consider the energy of a
system containing two vortices, it can be written
as $H \approx E(\theta^\ast) + E_0$, where $E_0$ is due to the creation of the
vortices at given positions, usually given as an integral over the whole
system. However, $E_0$ may be roughly approximated as
constant for a large system
since spins far away from the vortex pair will not be much affected by small
variations in their separation $\theta^\ast$.
Therefore, we expect that the dominant behavior to the total energy comes
from the interaction term, i.e., Eq.~(\ref{eq:u}), while $E_0$ only adds an
offset.

We can actually insert two vortices into this system, with one at the
origin and the other away from the origin.
A subtle part is that points at infinity, i.e., at
$|z|=1$, should be equally treated since there is no reason to distinguish
them. It is plausible that each field line should meet a point at
infinity at a right angle, as does a field line emitted from a defect at the
origin, since the exact vortex configuration near the origin will be
irrelevant at infinity.
This boundary-condition problem is
indeed equivalent to that of an electric charge inside a conducting cylinder
and one can solve this by introducing an image charge \emph{beyond}
infinity~\cite{furman}. That is, for a charge at $z_0$ with $|z_0|<1$,
its image charge should be located at $z_0' = \bar{z_0}^{-1}$ in order to
make every field line equally perpendicular to the boundary at infinity.
So we need $\gamma = \mbox{Im} \ln (z-z_0) - \mbox{Im} \ln (z-z_0')$
to insert the second vortex at $z=z_0$, and this leads to
\begin{equation}
\left| \frac{\partial \gamma}{\partial z} -A_z \right|^2 = \left|
\frac{1}{2i(z-z_0)} - \frac{1}{2i(z-z_0')} + \frac{1}{2iz} \left(
\frac{1+z\bar{z}}{1-z\bar{z}} \right) \right|^2.
\label{eq:two}
\end{equation}
Integrating this over a disk of radius $R<1$, we indeed find that $F \approx
-2 \pi K_A \theta^\ast + C_0$, where $\theta^\ast$ is the hyperbolic distance
between the vortices and $C_0$ is a system-dependent parameter
(Appendix~\ref{app:integral}). It agrees with the functional form in
Eq.~(\ref{eq:u}). The sign in front of $\theta^\ast$ is negative due to the
repulsive interaction since we are concerned with an analogy to the vortex
lattice in the theory of type-II superconductors.
To sum up, the interaction energy between two vortices is a linear function
of the hyperbolic distance between them unless they are very close.

\begin{figure}
\begin{center}
\includegraphics[width=0.23\textwidth]{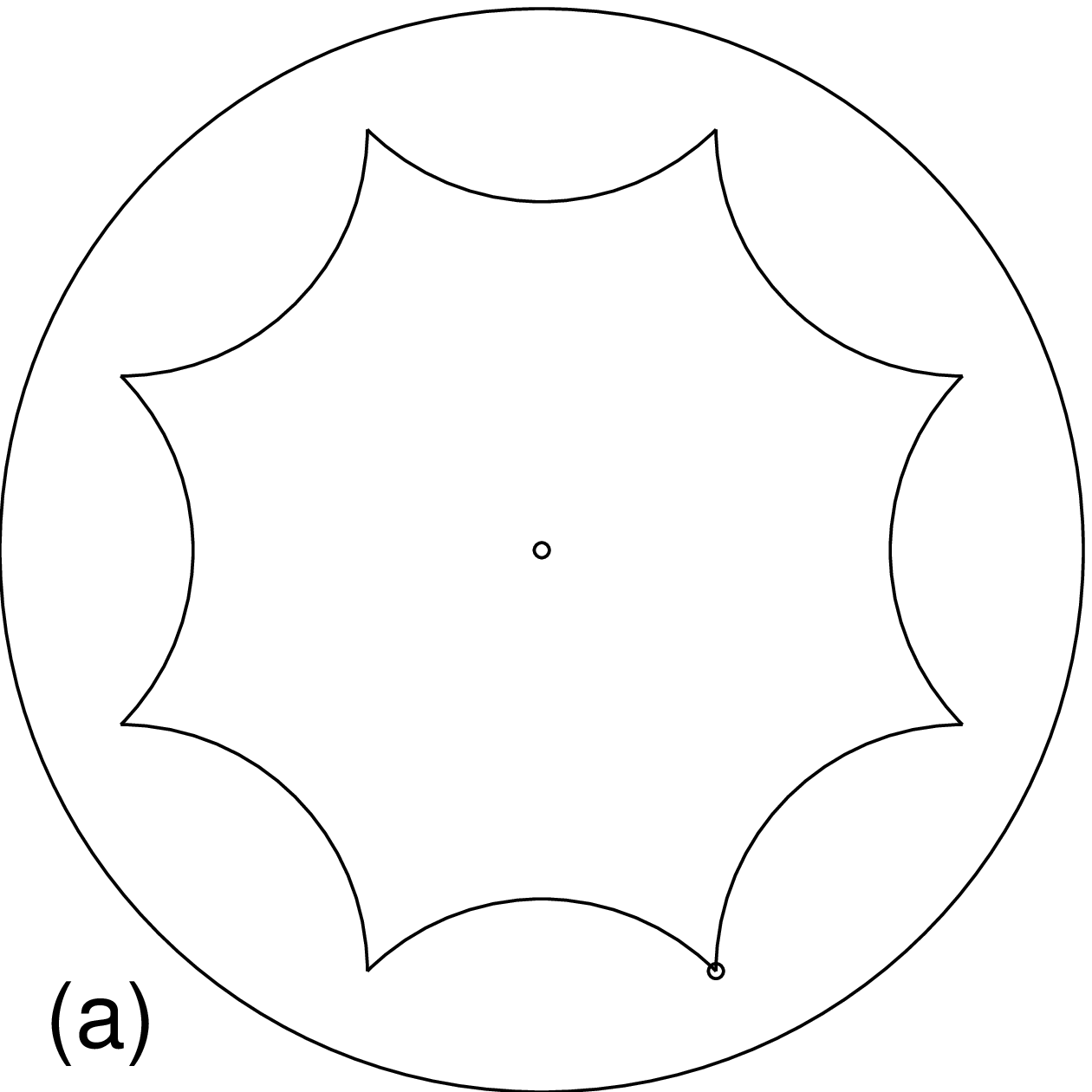}
\includegraphics[width=0.23\textwidth]{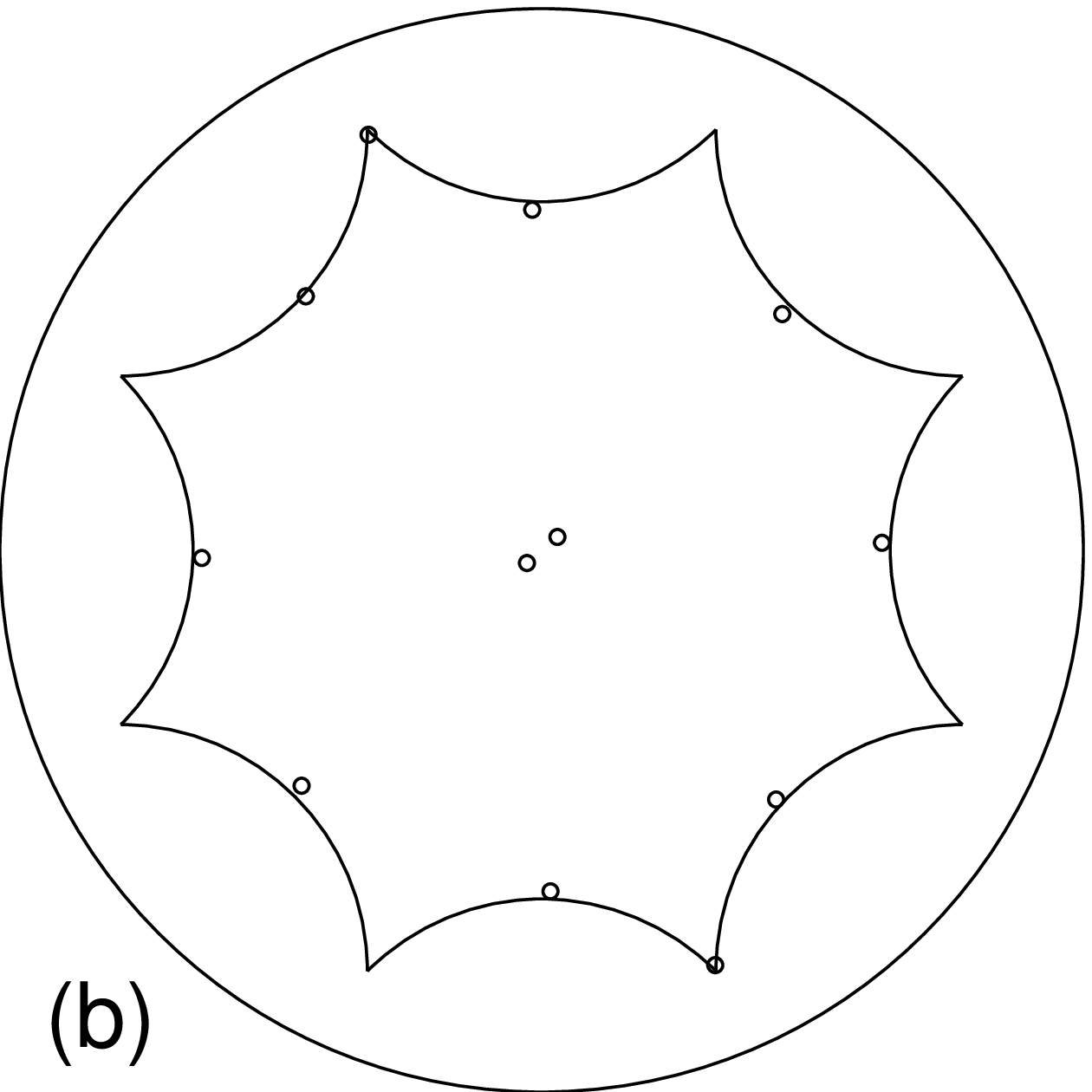}\\
\includegraphics[width=0.23\textwidth]{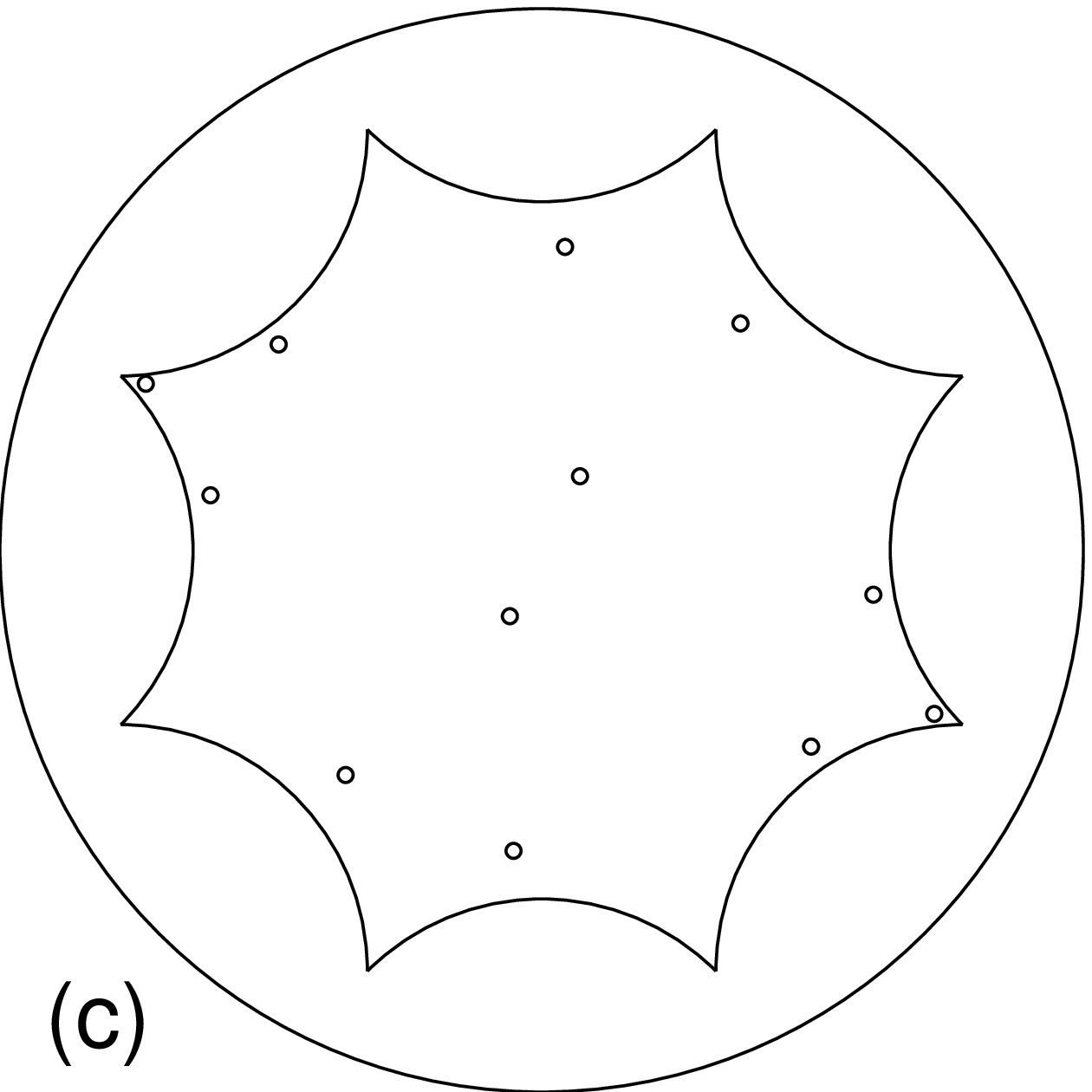}
\includegraphics[width=0.23\textwidth]{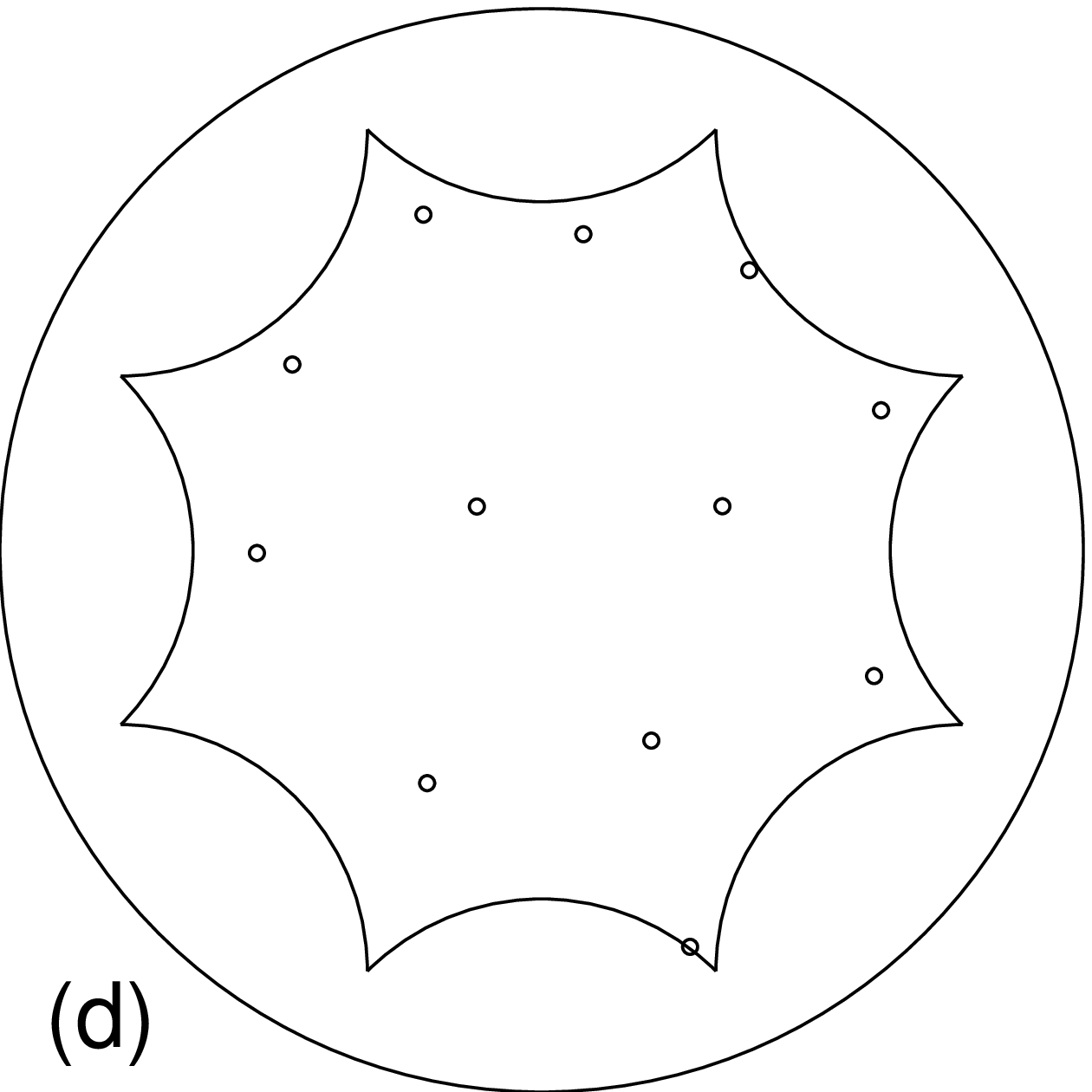}
\end{center}
\caption{Configurations of defects on a surface of a constant
negative curvature with the periodic boundary condition, represented on the
Poincar\'e disk. (a) When there exist two repulsive defects for
smetic-$C$ order, they are located at the maximum distance. (b) For hexatic
order, there are 12 defects interacting via the potential given by
Eq.~(\ref{eq:u}). (c) This pattern forms when the Coulomb potential
[Eq.~(\ref{eq:coulomb})] solely comes into play. (d) Every defect has seven
nearest neighbors when the potential is short ranged with a characteristic
hyperbolic distance.
}
\label{fig:thomson}
\end{figure}

Before concluding this work, we briefly consider a variant of the Thomson
problem, i.e., the Thomson problem on a pseudosphere. We can construct a
periodic boundary as suggested in Ref.~\cite{sausset} in such a way that we
merge every pair of opposite sides of the octagon in
Fig.~\ref{fig:thomson}. The resulting closed surface has genus $g=2$,
and therefore $\chi = 2-2g = -2$.
The distance from one defect to another is determined by the shortest one
among all the periodic images. Since this octagon is surrounded by 48
identical octagons, this means that we generally have 49 possible cases to
check for determining the distance.
Once the distance between every pair is found, it is straightforward to
compute the total energy by using a predefined potential function and
to find energy minima by applying the Metropolis algorithm.
In terms of liquid crystals, in the smetic-$C$ phase, the molecules
are tilted when measured relative to the surface normal and therefore
described by a usual vector field. Since $\chi = -2$, according to
Eq.~(\ref{eq:pb}), the surface should
have two defects, each of which has charge $-1$, or one dipole with charge
$-2$. The former case is depicted in Fig.~\ref{fig:thomson}(a), where we
find that the two repulsive defects are located at the largest possible
distance.
The hexatic phase, on the other hand, contains sixfold orientational order
and the surface has $12$ defects with charge $-1/6$ each~\cite{luben}. This
situation is shown in Fig.~\ref{fig:thomson}(b), where the defects
are observed as pairs. Up to this point, we have used the potential given
in Eq.~(\ref{eq:u}). But we may also ask ourselves what happens with
the potential shape is altered. For example, if only the Coulomb interaction
[Eq.~(\ref{eq:coulomb})] is present, a different pattern appears as shown in
Fig.~\ref{fig:thomson}(c).
In addition, Fig.~\ref{fig:thomson}(d) shows another case where the
interaction potential is assumed to have a certain `penetration depth'
$\lambda$, i.e., roughly given as $\exp(-\theta^{\ast}/\lambda)$ with
hyperbolic distance $\theta^{\ast}$ between a pair of defects.
In Fig.~\ref{fig:thomson}(d), the defects form a regular structure
where each of the 12 defects has a coordination number $k=7$ according to the
Euler-Poincar\'e relation and the Gauss-Bonnet theorem (see
Ref.~\cite{sausset10} for details).

\section{Summary}
\label{sec:summary}

In summary, we calculated the vortex-excitation energy on curved
surfaces by means of the stereographic projection. It was shown that the
interaction energy is a linear function of the distance between vortices
in the case of a negatively curved surface. This confirmed that the curvature
should appear in the source term of the generalized Gauss law. We also
explicitly obtained defect configurations minimizing the energy as a variant
of the Thomson problem.

\begin{acknowledgments}
The author is grateful to Jeong-Man Park, Petter Minnhagen, and Beom Jun Kim
for encouraging comments.
\end{acknowledgments}

%

\appendix

\section{Relation to two-dimensional electrostatics}
\label{app:elec}

Let us consider the Frank free energy, given by Eq.~(\ref{eq:f}), which has the
following form throughout this work:
\begin{equation}
F = 2 K_A \iint dz d\bar{z} \left( \frac{\partial \gamma}{\partial z} -
A_z \right) \left( \frac{\partial \gamma}{\partial z} - A_{\bar{z}} \right).
\label{eq:f2}
\end{equation}
For $\mathcal{L} \equiv  \left( \frac{\partial \gamma}{\partial z} -
A_z \right) \left( \frac{\partial \gamma}{\partial z} - A_{\bar{z}}
\right)$, the Euler-Lagrange equation with respect to $\gamma$ is written as
\begin{eqnarray}
0 &=&
\frac{\partial \mathcal{L}}{\partial \gamma}
- \frac{\partial}{\partial z} \frac{\partial \mathcal{L}}{\partial \left(
\frac{\partial \gamma}{\partial z} \right)} -
\frac{\partial}{\partial \bar{z}} \frac{\partial \mathcal{L}}{\partial
\left( \frac{\partial \gamma}{\partial \bar{z}} \right)}\nonumber\\
&=& 2\frac{\partial^2 \gamma}{\partial z \partial \bar{z}} - \frac{\partial
A_{\bar{z}}}{\partial z} - \frac{\partial A_z}{\partial \bar{z}}.
\label{eq:lagrange}
\end{eqnarray}
Substituting $A_z$ and $A_{\bar{z}}$ for the unit sphere
[Eq.~(\ref{eq:con1})] here, we see that the last two terms vanish
and we are left with
\begin{equation}
\frac{\partial^2 \gamma}{\partial z \partial \bar{z}} =
\frac{1}{4} \left( \frac{\partial^2}{\partial x^2} +
\frac{\partial^2}{\partial y^2} \right) \gamma(x,y) = 0,
\end{equation}
which is Laplace's equation in two dimensions. Therefore, for a given defect
configuration, the vector field satisfying the equivalent electrostatic
problem is the one that minimizes Eq.~(\ref{eq:f2}) over spin waves. The
same can be shown to be true on the Poincar\'e disk as well by substituting
Eq.~(\ref{eq:con2}) into Eq.~(\ref{eq:lagrange}).

\section{Integration of the first term in Eq.~(\ref{eq:frank})}
\label{app:term1}

Here we evaluate
$T_1 = \iint_D |z|^{-2} ~dz ~d\bar{z}$,
where $D$ is defined as a disk of $|z-z_0|<R$ containing the singularity at
the origin. First, we begin with Green's theorem:
\begin{equation*}
\oint_C L dx + M dy = \iint_D \left( \frac{\partial M}{\partial x} -
\frac{\partial L}{\partial y} \right) dx ~dy,
\end{equation*}
where $L$ and $M$ are functions of $(x,y)$ and $C$ is the boundary of $D$.
We want to find a vector field $\mathbf{F} = (F_x, F_y, F_z) = (L, M, F_z)$
such that 
$\left( \nabla \times \mathbf{F} \right)_z
= \frac{\partial M}{\partial x} - \frac{\partial L}{\partial y} =
(x^2 + y^2)^{-1}$.
Representing this in the cylindrical coordinate, we have
$\mathbf{F} = F_r \hat{\mathbf{r}} + F_{\phi} \hat{\mathbf{\phi}} + F_z
\hat{\mathbf{z}}$, and rewrite the above expression as
$\left( \nabla \times \mathbf{F} \right)_z
= \frac{1}{r} \left[ \frac{\partial}{\partial r} (r F_{\phi}) -
\frac{\partial F_r}{\partial \phi} \right] = r^{-2}$.
Letting $F_r = F_z = 0$, this yields
$\mathbf{F} = r^{-1} (\ln r + c) ~\hat{\mathbf{\phi}} = r^{-2} (\ln r
+c) \left( -r \sin\phi, r \cos\phi, 0 \right)$
with a constant $c$.
In addition, we have the following identity~\cite{mancill}:
\begin{equation*}
\oint_C (M + iL) ~dz = \oint_C (M~dx - L~dy) + i \oint_C (L~dx + M~dy).
\end{equation*}
The first part indeed vanishes for $L$ and $M$ above, since
\begin{eqnarray*}
\oint_C (M~dx - L~dy) &=& - \iint_D \left( \frac{\partial L}{\partial x} +
\frac{\partial M}{\partial y} \right) dx~dy\nonumber\\
&=& - \iint_D (\nabla \cdot \mathbf{F}) ~dx~dy = 0,
\end{eqnarray*}
which follows from $\nabla \cdot \mathbf{F} = \frac{1}{r}
\frac{\partial}{\partial r} (r F_r) +\frac{1}{r} \frac{\partial
F_{\phi}}{\partial \phi} + \frac{\partial F_z}{\partial z} = 0$.
Therefore, if we define a complex function $F(z) = L+iM$ as a counterpart of
$\mathbf{F}$, it leads to $M+iL = i \bar{F}$ and
\begin{equation*}
\oint_C \bar{F} dz = \oint_C (L~dx + M~dy) = \iint_D \left( \nabla \times
\mathbf{F} \right)_z ~dx~dy = \iint_D |z|^{-2} ~dz ~d\bar{z}.
\end{equation*}
Since
$\bar{F} = r^{-2} (\ln r +c) (-r \sin\phi  -ir \cos\phi) = r^{-2} (\ln r
+c) (y+ix) = r^{-2} (\ln r +c) ~i\bar{z}$,
the equation we are going to evaluate turns out to be
\begin{equation*}
\oint_C \bar{F} dz
= - \oint_C r^{-2} (\ln r +c) ~i\bar{z}~dz
= - \oint_C iz^{-1} \left( \frac{1}{2} \ln z\bar{z} +c \right) dz.
\end{equation*}
If the contour $C$ is given as $z\bar{z} = R^2$ by $z_0 = 0$, then the integral
becomes
$- \oint_C i z^{-1} \left( \ln R +c \right) dz = 2\pi (\ln R +c)$
by Cauchy's integral formula. Let us exclude a small disk $S$ of radius
$\delta \ll 1$ around the origin to remove the constant $c$. The integral on
the area between $D$ and $S$ is therefore $2\pi \ln \frac{R}{\delta}$.
In the case where $z_0$ is away from the origin, we assume that it is on the
positive real axis without loss of generality. In other words, we simply
have $z_0 = \rho$ with $0 < \rho < R-\delta$. The contour $C$ is now given
as $(z-\rho)(\bar{z}-\rho) = R^2$, or $\bar{z} = (\rho z + R^2 -
\rho^2)(z-\rho)^{-1}$. By inserting this, we get
\begin{equation*}
\oint_C \bar{F} dz = - \oint_C \frac{i}{z} \left[ \frac{1}{2} \ln z +
\frac{1}{2} \ln(\rho z + R^2 - \rho^2) - \frac{1}{2} \ln (z-\rho) +c \right]
dz \equiv G(\rho).
\end{equation*}
It is a bit cumbersome to directly evaluate the complex logarithms. We
alternatively differentiate it with $\rho$ and arrive at
\begin{equation}
\frac{\partial G}{\partial \rho} = -i \oint_C \left[
\frac{z-2\rho}{2z(\rho z + R^2 - \rho^2)} + \frac{1}{2z(z-\rho)} \right] dz.
\label{eq:dg}
\end{equation}
The first term has only one pole at $z=0$ while the second has two at $z=0$
and $z=\rho$, respectively. Applying Cauchy's integral formula once again,
it is found that
$\partial G / \partial \rho = - 2\pi \rho (R^2 - \rho^2)^{-1}$.
It is straightforward now to have
$G(\rho) = - \int 2\pi \rho (R^2 - \rho^2)^{-1} d\rho = \pi \ln \left(
1-\rho^2/R^2 \right) + c'$
with a new constant $c'$. However, we already know $G(\rho=0) = 2\pi (\ln R +
c)$ which determines $c'$. Furthermore, we exclude $S$ as before from the
integration range. The final result thus becomes
\begin{equation}
T_1 = 2 \pi \ln \frac{R}{\delta} + \pi \ln(1-\rho^2/R^2) = 2 \pi \ln
\frac{R}{\delta} \sqrt{1-\rho^2/R^2}.
\label{eq:t1}
\end{equation}
If $\rho = R-\delta$, for example, this formula yields $T_1 \approx \pi \ln
\frac{2R}{\delta}$, which can be cross checked by integrating
\[ \int_{\delta}^{2R} \frac{1}{r^2} 2 \phi r~dr
= \int_{\delta}^{2R} \frac{2}{r} \cos^{-1} \left( \frac{r}{2R} \right) dr
= \int_{\frac{\delta}{2R}}^{1} \frac{2}{y} \cos^{-1}y ~dy
\approx \pi \ln \frac{2R}{\delta}, \]
if we note that the circle centered at $(R,0)$ with radius $R$ is described
as $r = 2R \cos\phi$ in the $(r,\phi)$ coordinate.
As long as $R \gg \rho$, however, one can approximate Eq.~(\ref{eq:t1})
simply as $2 \pi \ln \frac{R}{\delta}$ and inserting $R = 4/\epsilon$ and
$\delta = \epsilon'$ gives the result in the main text.
What happens if $\rho>R$? Then the origin goes out of the contour
and $z=(\rho^2 - R^2)/\rho$ comes in instead since 
$\rho-R < (\rho^2 - R^2)/\rho < \rho+R$. It means that
Eq.~(\ref{eq:dg}) now yields
$\partial G / \partial \rho
= -2\pi (\rho^2+R^2)[2\rho (\rho^2 - R^2)]^{-1} + 2\pi (2\rho)^{-1}
= -2\pi R^2 [\rho (\rho^2 - R^2)]^{-1}$,
so we obtain
\begin{equation}
G(\rho) = -2\pi \int \frac{R^2}{\rho (\rho^2 - R^2)} d\rho = \pi \ln
\frac{1}{1-R^2/\rho^2}.
\label{eq:beyond}
\end{equation}
Note that the constant of integration is determined by making this
function vanish at $\rho \rightarrow \infty$.

\section{Integration of the last term in Eq.~(\ref{eq:frank})}
\label{app:term3}

In this appendix, we evaluate
\begin{equation*}
T_3 = \iint \frac{|z|^2 - (z \bar{z}_0 + \bar{z} z_0)/2}{
|z-z_0|^2 \left(1+|z|^2/4\right)} ~dz ~d\bar{z}.
\end{equation*}
By rewriting this using $z = r~e^{i\phi}$ and $z_0 = \rho$, we have
\begin{eqnarray*}
T_3 &=& \iint \frac{r^2 - r\rho \cos\phi}{r^2 + \rho^2 - 2r\rho\cos\phi}
\left( \frac{1}{1+r^2/4} \right) r dr d\phi\nonumber\\
&=& \int \frac{r~dr}{1+r^2/4} \int \frac{r^2 - r\rho
\cos\phi}{r^2 + \rho^2 - 2r\rho\cos\phi} d\phi.
\end{eqnarray*}
We first carry out the integration over $\phi$:
\begin{eqnarray*}
\int \frac{d\phi}{r^2 + \rho^2 - 2r\rho\cos\phi}
&=&\int \frac{d\phi}{r^2 + \rho^2 - r\rho \left( e^{i\phi} + e^{-i\phi}
\right)}\nonumber\\
&=&\int \frac{e^{i\phi} ~d\phi}{\left(r~e^{i\phi} - \rho \right)
\left(-\rho~e^{i\phi} +r \right)}\nonumber\\
&=& \frac{i}{\rho^2 - r^2} \int \left( \frac{1}{e^{i\phi} - \rho/r} -
\frac{1}{e^{i\phi} - r/\rho} \right) ie^{i\phi} ~d\phi\nonumber\\
&=& \frac{i}{\rho^2 - r^2} \oint \left( \frac{1}{\zeta - \rho/r} -
\frac{1}{\zeta - r/\rho} \right) d\zeta,
\end{eqnarray*}
with $\zeta = e^{i\phi}$. Note that we have a contour integral around a unit
circle centered at the origin. If $\rho/r<1$, then only the first term
contributes so we get $2\pi / (r^2-\rho^2)$ by Cauchy's integral formula. If
$r/\rho<1$, on the other hand, then only the second term contributes and we get
$2\pi / (\rho^2 - r^2)$. In short,
\begin{equation*}
\int \frac{d\phi}{r^2 + \rho^2 - 2r\rho\cos\phi}
= \frac{2\pi}{|r^2 - \rho^2|}.
\end{equation*}
Likewise,
\begin{eqnarray*}
\int \frac{\cos \phi ~d\phi}{r^2 + \rho^2 - 2r\rho\cos\phi}
&=&\int \frac{\frac{1}{2} \left( e^{i\phi} + e^{-i\phi} \right) d\phi}{r^2 +
\rho^2 - r\rho \left( e^{i\phi} + e^{-i\phi} \right)}\nonumber\\
&=&\frac{i}{2r\rho} \oint \frac{e^{i\phi} +
e^{-i\phi}}{(e^{i\phi}-\rho/r)(e^{i\phi}-r/\rho)} ie^{i\phi} d\phi \nonumber\\
&=&\frac{i}{2r\rho} \oint \frac{\zeta +
\zeta^{-1}}{(\zeta-\rho/r)(\zeta-r/\rho)} d\zeta\nonumber\\
&=&\frac{\pi}{r\rho} \left( \frac{r^2+\rho^2}{|r^2 - \rho^2|} - 1 \right).
\end{eqnarray*}
The integration over $\phi$ therefore yields
\begin{eqnarray*}
\int \frac{r^2 - r\rho \cos\phi}{r^2 + \rho^2 - 2r\rho \cos\phi} d\phi
&=& \pi \left( \frac{r^2-\rho^2}{|r^2-\rho^2|} +1 \right)\nonumber\\
&=& \left\{ \begin{array}{crl} 2\pi & \mbox{if} & r>\rho \\ 0 & \mbox{if} &
r<\rho \end{array} \right..
\end{eqnarray*}
Gathering the terms, we see that
\begin{equation*}
T_3 = 2\pi \int_{\rho}^{4/\epsilon} \frac{r ~dr}{1+r^2/4} = 8\pi
\left[ \ln (1+r^2/4) \right]_{r=\rho}^{4/\epsilon} \approx 4\pi \ln
\frac{4/\epsilon}{\epsilon \left(1 + \rho^2/4\right)},
\end{equation*}
which is the result in the main text.

\section{Integration of Eq.~(\ref{eq:two})}
\label{app:integral}

Again without loss of generality, we may set $z_0 = \rho$ with $0 < \rho < R
< 1$, where $R$ is the radius of the disk over which the integration
should be performed. Therefore, it follows that $z'_0 = \rho^{-1} \equiv
\rho' > 1$. The integrand [Eq.~(\ref{eq:two})] can be then written as
\begin{eqnarray*}
&&\left| \frac{1}{2i(z-\rho)} - \frac{1}{2i(z-\rho')} + \frac{1}{2iz} \left(
\frac{1+z\bar{z}}{1-z\bar{z}} \right) \right|^2\\
&=& \frac{1}{4\left| z-\rho \right|^2}
+ \frac{1}{4\left| z-\rho' \right|^2}
+ \frac{1}{4|z|^2} \left( \frac{1+|z|^2}{1-|z|^2}\right)^2\\
&-& \frac{1}{4(z-\rho)(\bar{z}-\rho')}
- \frac{1}{4(\bar{z}-\rho)(z-\rho')}\\
&+& \frac{1}{4\bar{z}(z-\rho)} \left( \frac{1+|z|^2}{1-|z|^2} \right)
+ \frac{1}{4z(\bar{z}-\rho)} \left( \frac{1+|z|^2}{1-|z|^2} \right)\\
&-& \frac{1}{4\bar{z}(z-\rho')} \left( \frac{1+|z|^2}{1-|z|^2} \right)
- \frac{1}{4z(\bar{z}-\rho')} \left( \frac{1+|z|^2}{1-|z|^2} \right).
\end{eqnarray*}
The integration of the first term has been already done in
Appendix~\ref{app:term1}:
\[ \iint_{|z|<R} \frac{dz d\bar{z}}{4\left| z-\rho \right|^2} = \frac{\pi}{2}
\ln \frac{R}{\delta} \sqrt{1-\rho^2/R^2}, \]
where $\delta \ll 1$ is the radius of a small circle around $z_0$ to be
excluded from the integration.
We have also obtained the result for the second term:
\[ \iint_{|z|<R} \frac{dz d\bar{z}}{4\left| z-\rho' \right|^2}
= \frac{\pi}{4} \ln \frac{1}{1-R^2/ \rho'^2}
= \frac{\pi}{2} \ln \frac{1}{\sqrt{1-\rho^2 R^2}} \]
since $\rho' > 1$ [Eq.~(\ref{eq:beyond})]. If $R \approx 1$, the
contribution from these two terms will be approximately $\frac{\pi}{2} \ln
\frac{R}{\delta}$, losing the dependence on $\rho$.
By representing $z = re^{i\phi}$, we see that the third term is a function of
$r$ only and can be integrated directly. However, the important point is
that the result cannot have any dependence on $\rho$.
\begin{eqnarray*}
\int_\delta^R \frac{1}{4r^2} \left( \frac{1+r^2}{1-r^2} \right)^2 2\pi r dr
&=& \left. \frac{\pi}{2} \left( \ln r + \frac{2}{1 - r^2} \right)
\right|_{r=\delta}^R \\
&=& \frac{\pi}{2} \left( \ln \frac{R}{\delta} + \frac{2}{1 - R^2} -
\frac{2}{1 - \delta^2} \right) \\
&\approx& \frac{\pi}{2} \left( \ln\frac{R}{\delta} + \frac{2R^2}{1-R^2}
\right).
\end{eqnarray*}
The fourth and fifth terms vanish together, which can be shown by a direct
integration as follows:
\begin{eqnarray*}
&& \iint_{|z|<R} \frac{1}{2} \left[ \frac{1}{(z-\rho)(\bar{z}-\rho')}
+\frac{1}{(\bar{z}-\rho)(z-\rho')} \right] dz d\bar{z}\\
&=& \int^R r~dr \int_{-\pi}^{\pi} d\phi \frac{r^2+1-r\cos\phi
(\rho+\rho')}{\left[ r^2+1 - r\cos\phi(\rho + \rho')\right]^2 +
r^2\sin^2\phi (\rho - \rho')^2}\\
&=& \int^R r~dr \left\{ \frac{1}{1-r^2} \arctan \left[
\frac{(1-r^2)\sin\phi}{r^2 \cos\phi - r(\rho+\rho') + \cos\phi} \right]
\right\}_{\phi=-\pi}^{\pi} = 0.
\end{eqnarray*}
We proceed to the sixth and seventh terms. It is these terms that are the
most relevant in this calculation since they describe the interaction
between the two defects, with one at the origin and the other at $z_0 = \rho$.
Together, they can be expressed as 
\[ \frac{1}{4} \left[ \frac{1}{\bar{z}(z-\rho)} + \frac{1}{z(\bar{z}-\rho)}
\right] \left( \frac{1+r^2}{1-r^2} \right)
= \frac{1}{2} \left(
\frac{1 - \frac{\rho}{r} \cos\phi}{r^2 + \rho^2 - 2r \rho \cos\phi } \right)
 \left( \frac{1+r^2}{1-r^2} \right). \]
Recalling Appendix~\ref{app:term3}, we notice that
\begin{eqnarray}
\int_{-\pi}^{\pi}
\frac{1 - \frac{\rho}{r} \cos\phi}{r^2 + \rho^2 - 2r \rho \cos\phi} d\phi
&=& \frac{2\pi}{|r^2 - \rho^2|} - \frac{\pi}{r^2} \left(
\frac{r^2+\rho^2}{|r^2 - \rho^2|} -1 \right)\nonumber\\
&=& \left\{ \begin{array}{ll}
2 \pi /r^2 & \mbox{if~}r > \rho\\
0 & \mbox{if~}r<\rho.
\end{array} \right.
\label{eq:van}
\end{eqnarray}
Therefore, the remaining integration of the sixth and seventh terms over $r$
corresponds to
\begin{eqnarray}
\pi \int_\rho^R \frac{1}{r^2} \left( \frac{1+r^2}{1-r^2} \right) r~dr
&=& \pi \left[ -r + 2 \tanh^{-1} r \right]_{r=\rho}^R \nonumber\\
&=& \pi \left[ -R + 2 \tanh^{-1} R + \rho - 2 \tanh^{-1} \rho \right]
\label{eq:final}
\end{eqnarray}
Lastly, the eighth and ninth terms vanish according to
Eq.~(\ref{eq:van}) since $r$ is always smaller than $\rho' > 1$.

To sum up, the $\rho$ dependence essentially originates from
Eq.~(\ref{eq:final}). By transforming $\rho$ to the corresponding
hyperbolic distance $\theta^\ast = 2 \tanh^{-1} \rho$, we therefore conclude
that the Frank free energy from the integration of Eq.~(\ref{eq:two})
asymptotically results in
\[ F = 2 K_A \int_{|z|<R} \left| \frac{\partial \gamma}{\partial
z} - A_z \right|^2 dz d\bar{z} \approx - 2 \pi K_A \theta^\ast + C_0 (R,
\delta), \]
where $C_0(R, \delta)$ is a system-dependent parameter.

\end{document}